\documentclass[aps,pre,twocolumn,byrevtex,showpacs]{revtex4}

\def\ogonek#1{\setbox0=\hbox{`}\ooalign{%
  \hidewidth\lower\ht0\copy0\hidewidth\crcr#1\crcr}}

\usepackage{epsfig}

\usepackage[latin1]{inputenc}
\usepackage[english]{babel}

\def\be{\begin{equation}} \def\eea{\end{eqnarray}}
\def\ee{\end{equation}} \def\bea{\begin{eqnarray}}
\def\ea{\end{array}} \def\ba{\begin{array}}

\newcommand{\bel}[1]{\begin{equation}\label{#1}}

\def\zzz{{\mathchoice {\hbox{$\sf\textstyle Z\kern-0.4em Z$}}
{\hbox{$\sf\scriptstyle Z\kern-0.3em Z$}}
{\hbox{$\sf\scriptscriptstyle Z\kern-0.2em Z$}} {\hbox{$\sf\textstyle
Z\kern-0.4em Z$}}}}

\begin{document}

%\title{Complete phase diagram for amnestically induced persistence}

\begin{abstract}
%We derive from first principles a linearizable expression for the
%nonlinear photocurrent-voltage characteristics of nanocrystalline
%TiO$_2$ dye sensitized solar cells.
We propose a new linearizable model for the nonlinear
photocurrent-voltage characteristics of nanocrystalline TiO$_2$ dye
sensitized solar cells based on first principles
and report predicted values for fill factors.
Upon renormalization
diverse experimental photocurrent-voltage data collapse onto a single
universal function.
%This result allows the estimation of the
%complete current-voltage curve from any three experimental
%measurements --- for instance the open circuit voltage, the short
%circuit current and the fill factor.
These advances allow the estimation of the complete current-voltage
curve and the fill factor from any three experimental data points,
e.g., the open circuit voltage, the short circuit current and one
intermediate measurement.
%and reinterpret existing
%data as evidence that the concentration of the oxidized redox species
%scales with the electron population in the semiconductor with a
%fractional power law exponent.
The theoretical underpinning provides
insight into the physical mechanisms responsible for the remarkably
large fill factors
% of dye sensitized solar cells,
as well as their
known dependence on the open circuit voltage.
%
%
%allow us to reinterpret existing experimental data as indicative of a
%reduction of the concentration of electron acceptor redox species
%I$_3^-$ with increasing voltage.

\end{abstract}

\title{Universal photocurrent-voltage characteristics of\\
dye sensitized nanocrystalline TiO$_2$ photoelectrochemical cells}

\author{J. S. Agnaldo} \author{J. C. Cressoni} \author
{G. M. Viswanathan} \affiliation{Laboratório de Energia Solar,
Núcleo de Tecnologias e Sistemas Complexos, \\
Instituto de F\'{\i}sica, Universidade Federal de Alagoas,
Macei\'{o}--AL, 57072-970, Brazil}

\date{\today}

\maketitle

\section{Introduction}

An important feature of photovoltaic solar cells and of diverse
optoelectronic devices studied in semiconductor physics concerns their
current-voltage characteristics~\cite{livro,livro2,livro3,prl-iv}. The
pioneering work that led to the invention of Grätzel or dye sensitized
solar cells became a milestone in the study of photovoltaic
devices~\cite{gratzel,gratzel2}. Previous theoretical and experimental
studies have identified the dependence of the photocurrent and
photovoltage on radiant power~\cite{huang}, but not the precise
nonlinear dependence of the photocurrent on the photovoltage under
conditions of constant radiant power. Moreover, variability in the
manufacturing process of dye sensitized solar cells can lead to
differences --- e.g., variables include the choice of dye, the
sintering temperature, thickness of the nanoporous TiO$_2$ film and
choice of chemical treatments. This diversity leads to significant
qualitative and quantitative variation in photocurrent-voltage
characteristics and of the relevant quantities such as the open
circuit voltage $V_{\mbox{\scriptsize \scriptsize oc}}$ or the fill
factor.  Such variability has discouraged attempts to identify
(possibly ``hidden'') dynamical patterns that could yield important
insights into the regenerative photoelectrochemical mechanisms that
underlie the conversion process.  Given the variability and diversity
in the characteristics, which properties remain universal and which
nonuniversal?  More importantly from an experimental point of view,
how can we quantitatively model the photocurrent-voltage
characteristics, based on fundamental theoretical principles?  Here we
answer these questions by deriving from first principles an analytical
expression for the photocurrent.

The topic of solar energy in
general~\cite{solar-general,solar-general2,solar-general3,solar-general4}
and dye sensitized solar cells in
particular~\cite{gratzel,gratzel2,huang,gratzel3,deb,durrant,frank1,frank2,pico,nelson}
attracts broad interest from diverse sectors of society, due to
technological, economic, political and environmental
considerations. The growing scientific interest in dye sensitized
TiO$_2$ solar cells stems from their unusual features and mode of
operation that distinguish them from Si solar cells: (i) efficient
charge separation due to ultra-fast injection of electron from the
dye on picosecond and subpicosecond time
scales~\cite{durrant,pico}; (ii) conduction consisting only of
injected electrons rather than electron-hole
pairs~\cite{gratzel,gratzel2}, due to the wide bandgap of the
semiconductor TiO$_2$; (ii) high optical density due to the
extremely large surface area of the dye sensitized nanoporous
semiconductor~\cite{durrant}; (iv) negligible charge recombination
with the oxidized sensitizer dye~\cite{gratzel,gratzel2,huang};
and
(v) very high quantum yields~\cite{durrant}.
One important fact that will contribute towards derivation and
subsequent interpretation of the current-voltage characteristics
concerns how the experimentally measured recombination current density
vanishes at short circuit~\cite{huang} --- indicating that the only
significant recombination pathway proceeds via back electron transfer
into the electrolyte.  Indeed, charge recombination between redox
species (I$_3^-$ ions) in the electrolyte and conduction band
electrons localized at the nanoporous interface result in suboptimal
photovoltage levels --- thus limiting the conversion
efficiency~\cite{suboptmal,suboptmal3,suboptmal4}.
%~\cite{huang's 20,21-23}.

\begin{figure}[t]
\centerline{\psfig{width=8 cm,clip,figure=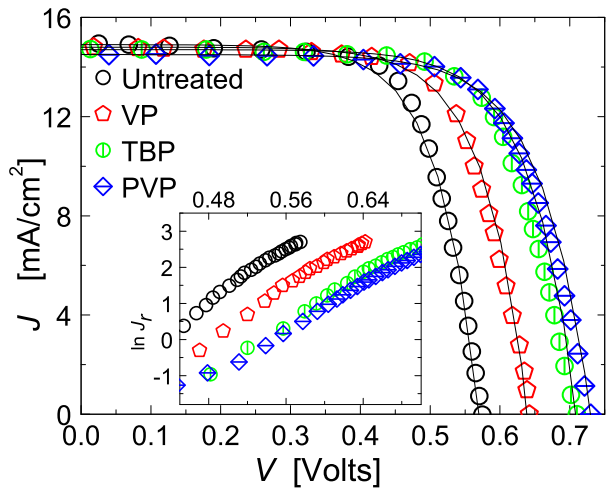}~~~}
\caption{Typical photocurrent voltage characteristics under a radiant
power of 1.5 AM for dye sensitized solar cells, taken from
ref.~\protect\cite{huang}.  The four data sets have different
characteristics due to varied chemical treatments (see text), yet in
each case the theoretical curves (solid lines) corresponding to
Eq.~\ref{eq-main} can account well for the experimental curves.  Inset
shows approximate logarithmic relation for the recombination current
density $J_{\mbox{\scriptsize r}} $ versus voltage, not inconsistent with Eq.~\ref{eq-jr}.}
\label{fig-iv}
\end{figure}

\begin{figure}[t]
\centerline{\psfig{width=8 cm,clip,figure=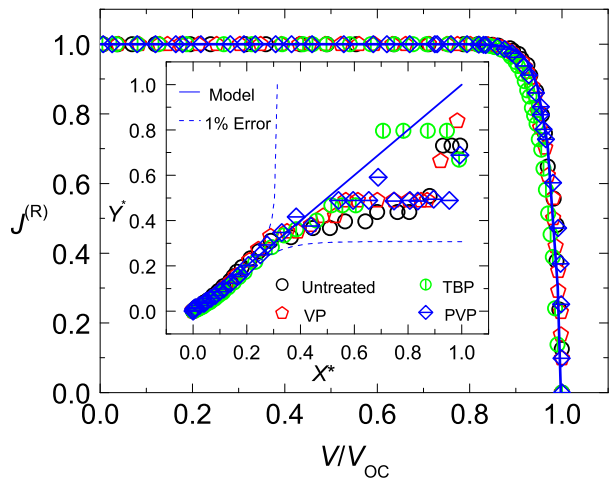}~~~}
\caption{Renormalized photocurrent $J^{(R)}$ versus
$V/V_{\mbox{\scriptsize oc}}$ for the curves shown in
Fig.~\ref{fig-iv}. The data collapse onto a single universal
function stable over a variety of chemical treatments and solar
cell variability. We have chosen $v_{\mbox{\scriptsize s}}=1/40$,
based on the value of $k_{\mbox{\tiny B}}T$ at room temperature, for
illustration, corresponding to our estimate of an upper bound for
fill factors, however we could have renormalized the photocurrent
to any fill factor. Indeed, we can linearize the curves, as shown
in the inset and explained in the text. The dashed line traces the
upper and lower uncertainties corresponding to an error of 1\% in
the short circuit current. Notice how remarkably the data collapse
onto a straight line, within the error tolerance.} \label{fig-dc}
\end{figure}

\begin{figure}[t]
\centerline{\psfig{width=8.5 cm,clip,figure=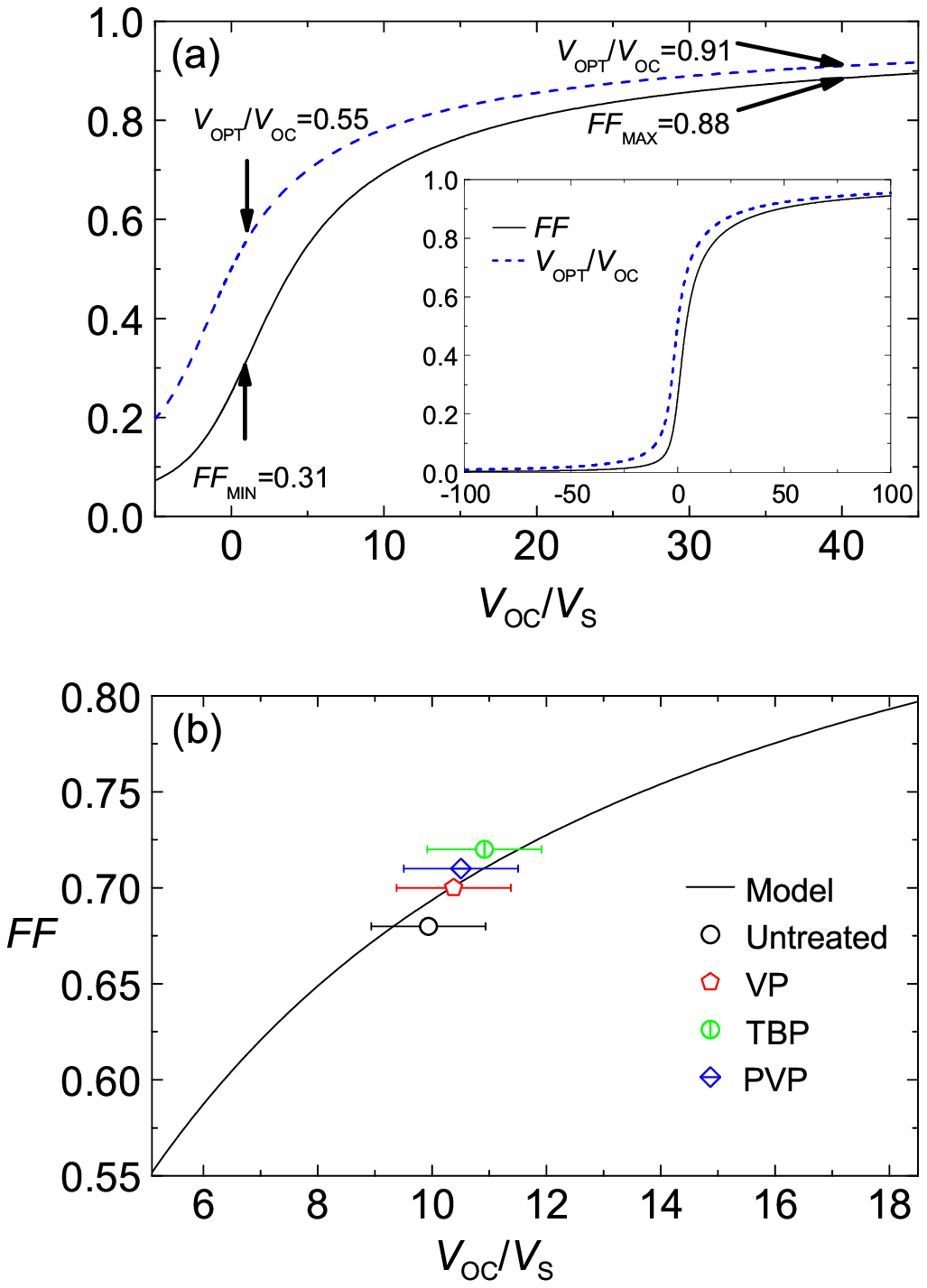}~~~}
\caption{ (a) Theoretically predicted fill factors FF and ratio
$V_{\mbox{\tiny \tiny OPT}}/V_{\mbox{\scriptsize oc}}$ of optimal
operating voltage $V_{\mbox{\scriptsize \tiny OPT}}$ to open circuit
voltage $V_{\mbox{\scriptsize oc}}$ versus $V_{\mbox{\scriptsize
oc}}/V_{\mbox{\scriptsize s}}$, where $V_{\mbox{\scriptsize s}}$
represents a characteristic voltage (Eq.~\protect\ref{eq-vs}).  We
estimate a worst-case lower bound on the fill factor using
$  V_{\mbox{\scriptsize
oc}}/ V_{\mbox{\scriptsize s}}=1$, and an idealized upper bound using
$V_{\mbox{\scriptsize
oc}}/V_{\mbox{\scriptsize s}}=40$.  Inset shows a more complete
picture. (b) Comparison of theoretical and experimental values for FF
for the data shown in Fig~\protect\ref{fig-iv}. Useful real cells will
likely have FF within the range shown.}

\label{fig-ff}
\end{figure}

\section{Methods}

We begin by assuming that the number of electrons injected into the
conduction band depends only on the incident radiant power --- in fact
the known very high quantum yields justify this assumption.  This
assumption allows us to express the recombination current density
$J_{\mbox{\scriptsize r}} $ as a function of the photocurrent density
$J$ and the injection current density. Since $J_{\mbox{\scriptsize r}}
$ vanishes as short circuit, the injection current equals the short
circuit current $J_{\mbox{\scriptsize sc}}$, so that
\bel{eq-1}
J_{\mbox{\scriptsize r}} =J_{\mbox{\scriptsize sc}}-J \;\; .
\ee
For a well mixed solution with identical surface and bulk
concentrations (typical for small current densities), the
Butler-Volmer equation~\cite{bv1,bv2} leads to the following
expression for the recombination current density $J_{\mbox{\scriptsize r}} $:
\bel{eq-bv}
-J_{\mbox{\scriptsize r}} =J_0 \bigg[
\exp(-{\alpha_{\mbox{\tiny C}} uf}\eta)-
\exp({\alpha_{\mbox{\tiny A}} uf}\eta)
\bigg]\;\ ,
\ee
%
%link:
%http://de.wikipedia.org/wiki/Butler-Volmer-Gleichung
%
%http://209.85.165.104/search?q=cache:Ppy1DV9uH7wJ:www.cartage.org.lb/en/themes/Sciences/Chemistry/Electrochemis/Electrochemical/ElectrodeKinetics/ElectrodeKinetics.htm+butler-volmer-equation&hl=pt-BR&ct=clnk&cd=4&gl=br
%
%
where $J_0$ denotes the exchange current density, $u$
the number of electrons transferred in the reaction (and consequently,
 the order of the rate of reaction for recombination for electrons)
$\alpha_{\mbox{\tiny A}}$ and $\alpha_{\mbox{\tiny C}}$ the anodic and cathodic transfer
coefficients, $\eta=V$ the overpotential and $f\equiv q/k_{\mbox{\tiny B}}T$.  With
some simplification~\cite{huang}, this equation becomes
\bel{eq-jr} J_{\mbox{\scriptsize r}} =qk_{\mbox{\scriptsize et}}c^m
n_0^{u\alpha}[\exp(u\alpha qV/k_{\mbox{\tiny B}}T)-1]\;\;, \ee
where $\alpha=\alpha_{\mbox{\tiny A}}$ and $k_{\mbox{\scriptsize et}}$
denotes the back electron transfer rate constant, $c$ the
concentration and $m$ the order
of the reaction
for the oxidized species, $q$
the electronic  charge and $n_0$ represents value in dark conditions of the
electron population in the semiconductor.

How do the prefactors depend on voltage $V$?  The back electron
transfer rate constant varies with radiant power and the maximum,
i.e. open circuit, photovoltage $V_{\mbox{\scriptsize oc}}$,
however we do not expect it to depend on the photovoltage for
fixed radiant power.
%On the other hand, we know that the redox
%species concentration in the electrolyte varies along the space
%     between the two electrodes, across which the voltage drops during
%operation.
The Nernst Equation
%$ \Delta V= -({k_{\mbox{\tiny B}}T}/ q) \ln
%({c_{ox}}/{c_{red}}),$
for the potential in terms of
%the standard electrode potential $E^0$
the concentrations of the oxidized and reduced species
%$c_{ox}$ and
%$c_{red}$
%respectively
holds valid only under equilibrium conditions,
yet we know that $c$ varies across the electrolyte.  Indeed,
since the reduced species greatly exceeds the oxidized species, we can
safely conclude that the voltage varies as $\Delta V(x) \approx
-(k_{\mbox{\tiny B}}T/q)\ln(c^{0}/c(x))$, where $c(x)$ denotes the
concentration at a position $x$ across the cell (i.e., electrolyte),
$c^0$ denotes a reference (or mean) concentration. Nonetheless, we
still do not know, {\it a priori}, exactly how it varies with the
potential at the semiconductor-electrolyte interface as the external
load (i.e., impedance) varies, because of the out of equilibrium
conditions.  In this context, one important clue comes from the
dependence on the photovoltage on the electron population.  Electrons
act as charge carriers in the TiO$_2$ --- just as the redox species do
in the electrolyte.  In the semiconductor, injected electrons shift
the Fermi level, so that $n=n_0 \exp(qV/k_{\mbox{\tiny B}}T)$.  This
exponential dependence on potential, together with the exponential
dependence of $c$ on voltage across the electrolyte,
hints at
a similar, i.e. exponential, dependence of $c$ on $V$ at the
interface as the external load varies:
\bel{eq-nova}
 V=V_{\mbox{\scriptsize oc}} -(\gamma k_{\mbox{\tiny
B}}T/q) \ln({{c_{\mbox{\scriptsize oc}}}/{c}})\;\;.
\ee
Here $c_{\mbox{\scriptsize oc}}$ represents the concentration under
open circuit conditions and $\gamma $ represents a free parameter in
the model, quantifying the fraction of the voltage variation that
affects the oxidized species concentration.  Since the triiodide
concentration cannot vary very much in the liquid, we cannot expect
small $\gamma $ close to equal unity since this would imply $c\propto
n$.  For now we only mention that one would naïvely expect a positive
$\gamma$ since electron injection and dye regeneration (associated
with larger $V$) produce oxidized species.

\section{Results}

These considerations immediately lead to an analytical expression for
the photocurrent $J$ as a function of the voltage $V$ across the cell:
\bea
\label{eq-full} 
J= \nonumber J_{\mbox{\scriptsize sc}}
\bigg[ 1-
\frac{\exp(mq(V-V_{\mbox{\scriptsize oc}})/\gamma k_{\mbox{\tiny
B}}T)} {\exp(u\alpha qV_{\mbox{\scriptsize oc}}/k_{\mbox{\tiny B}}T
)-1} \times \\
\bigg( \exp(u\alpha q V/k_{\mbox{\tiny B}}T) -1 \bigg) \bigg]
\;\; . 
\eea 
Here $m\approx 2$ because of the second order reaction and
$u\alpha\approx 0.7$ ~\cite{huang,deb}.  This analytical expression
must hold true for all dye sensitized solar cells, yet in the context
of providing greater clarity and insight we can render it simpler but
more useful. Moreover, from a practical point of view, we can simplify
it further by making additional yet realistic
assumptions. Eq. \ref{eq-jr} for the recombination has validity in the
large voltage ($V>80$mV) regime~\cite{huang,deb}. On the other hand,
below this potential, the recombination current becomes negligible and
uninteresting.  For any useful cell, $n\gg n_0$ by many orders of
magnitude, so that we can reasonably well approximate
Eq. \ref{eq-full} with
\bel{eq-main}
J=J_{\mbox{\scriptsize sc}}\left[
\frac
{\exp( V_{\mbox{\scriptsize oc}}/V_{\mbox{\scriptsize s}})  - \exp( V/V_{\mbox{\scriptsize s}}) }
{\exp( V_{\mbox{\scriptsize oc}}/V_{\mbox{\scriptsize s}})  - 1}
\right] \;\; ,
\ee
where
the potential
\bel{eq-vs} V_{\mbox{\scriptsize s}}\equiv (k_{\mbox{\tiny B}}T/q) \frac
1{u\alpha+m/\gamma }\approx \frac1{40(0.7+2/\gamma ) } \rm Volts
\;\; , \ee
represents a characteristic scale of the exponential decay.
Specifically, $V_{\mbox{\scriptsize s}}$ quantifies the photovoltage
drop corresponding to a decrease in recombination current density by a
factor of $1/e$ where $e$ here denotes Euler's number.
Eqs. \ref{eq-full} and \ref{eq-main} represent the first out of three
new results of this article.
%\end{document}

Fig. \ref{fig-iv} compares the model with photocurrent-voltage curves
taken from ref.~\cite{huang}, of untreated and pyridine
derivative-treated [RuL$_2$(NCL)$_2$]-coated nanocrystalline TiO$_2$
electrodes in CH$_3$CN/MNO (50:50 wt \%) containing Li(0.3M) and
I$_2$(30mM), for a radiant power of 100mW/cm$^2$ (AM 1.5).  The
electrodes had treatment with following substances: 3-vinylpyridine
(VP), 4-$tert$-butylpyridine (TBP), and poly(2-vinylpyridine) (PVP).
The good agreement with the data validates the model represented by
Eqs. \ref{eq-full} and \ref{eq-main}.

The largest possible power output divided by $J_{\mbox{\scriptsize
sc}} V_{\mbox{\scriptsize oc}}$ defines the fill factor FF. Notice
that $V_{\mbox{\scriptsize s}}$ changes the fill factor (via $\gamma
$). A value $V_{\mbox{\scriptsize s}}\rightarrow \infty$
(corresponding to purely resistive or Ohmic behavior) leads to FF=1/4,
whereas $V_{\mbox{\scriptsize s}}\rightarrow 0$ leads to unity fill
factor --- perfect but theoretically impossible except at
$T=0$~K. Most dye sensitized solar cells have FF=0.6--0.7
(Fig.~\ref{fig-iv}).

We now turn our attention to the question of whether a single
universal current-voltage relation can describe all TiO$_2$ solar
cells.  According to the theory presented above, all dye sensitized
solar cells must satisfy Eq.~\ref{eq-full} if not
Eq. ~\ref{eq-main}. If we renormalize the photovoltage to obtain an
adimensional measure $V^*\equiv V/V_{\mbox{\scriptsize oc}}$, then
every single dye sensitized solar cell must satisfy the following
relation for an idealized renormalized photocurrent:
\bel{eq-ideal}
J^{(R)}\equiv
\frac
{1-\bigg[1-(J/J_{\mbox{\scriptsize sc}})\bigg(1-\exp(-V_{\mbox{\scriptsize oc}}/V_{\mbox{\scriptsize s}})\bigg)\bigg]  ^{ V_{\mbox{\scriptsize s}} /V_{\mbox{\scriptsize oc}} v_{\mbox{\scriptsize s}}    }  }
{1-\exp(-1/v_{\mbox{\scriptsize s}})}      \;\;\; .
\ee
Here $v_{\mbox{\scriptsize s}}$ fixes the shape or fill factor of the renormalized
photocurrent.  Fig. \ref{fig-dc} shows the predicted data collapse.
We have chosen a value $v_{\mbox{\scriptsize s}}=1/40$, due to its significance for an
idealized solar cell with maximum FF (see below) at room
temperature. However, we can obtain data collapse for any $v_{\mbox{\scriptsize s}}$ (not
shown). This is perhaps more clear if we linearize the curves. We
define coordinates
\bea
X^*&\equiv& 1-V/V_{\mbox{\scriptsize oc}}  \\
Y^*&\equiv&
-V_{\mbox{\scriptsize s}}/V_{\mbox{\scriptsize oc}} \ln\bigg[ 1-\frac J {J_{\mbox{\scriptsize sc}}} \bigg(1-\exp(-V_{\mbox{\scriptsize oc}}/V_{\mbox{\scriptsize s}})  \bigg)\bigg] \;\;.
\label{eq-xy}
\eea
The inset of Fig.~\ref{fig-dc} shows how the data collapse onto a
straight line.
All dye sensitized solar cells thus follow the same universal
pattern of photocurrent-voltage behavior.

We next consider the problem from the point of view of scale
invariance symmetry.  The fill factor cannot depend on
$J_{\mbox{\scriptsize sc}}$, since it cancels in the power ratio. It
also remains invariant under a scale transformation
%
%\be
\mbox{$ V_{\mbox{\scriptsize oc}} \rightarrow \lambda
V_{\mbox{\scriptsize oc}}, ~V_{s} \rightarrow \lambda V_{s}.$}
%\ee
%
In fact, no dilation can alter a ratio of geometric areas. The
invariance of FF for arbitrary $\lambda$ implies that FF can depend on
$V_{\mbox{\scriptsize oc}}$ and $V_{\mbox{\scriptsize s}}$ only via
their ratio:
\be
FF=FF(V_{\mbox{\scriptsize oc}}/V_{\mbox{\scriptsize s}})   \;\; .
\ee
The exact functional dependence appears to involve a transcendental
equation.  We are still attempting an analytical solution using the
Lambert W function.  Nevertheless, it is susceptible to numerical
solution.  Fig.~\ref{fig-ff}(a) shows FF as a function of
$V_{\mbox{\scriptsize oc}}/V_{\mbox{\scriptsize s}}$.

We next comment on the values typically found for
$V_{\mbox{\scriptsize s}}$ and their physical significance. The values
found correspond to negative $\gamma $ and thus suggest that the
concentration of redox species (I$_3^-$ ions) decreases with the
photovoltage. This may at first seem counter-intuitive.  Indeed,
higher voltage suggests larger electron population and more injection.
Moreover, the regeneration of the dye creates I$_3^-$ species, in the
proportion of one ion for every two electrons injected.

So do we face an apparent inconsistency?  The important fact,
mentioned earlier, of zero recombination current density $J_{\mbox{\scriptsize r}} =0$ under
short circuit conditions, hints at the correct explanation: the rate
of regeneration of the oxidized dye depends not on the photovoltage
(zero under short circuit) but rather on the rate of electron
injection --- thus only on the open circuit photovoltage, or
alternatively, on the radiant power.
The finding
agrees with the expectation of a smaller depletion layer for more
external current drain.

The above findings allow us to estimate lower and upper limits for
FF. Purely Ohmic behavior corresponds to $V_{\mbox{\scriptsize
s}}\rightarrow \infty$ and FF=1/4, however we cannot imagine this
scenario.  For any useful device, the largest conceivable value of
$V_{\mbox{\scriptsize s}}$ should not exceed $V_{\mbox{\scriptsize
oc}}$, which gives us a lower bound for FF of FF=0.31 and an optimal
operational voltage of $V_{\mbox{\scriptsize \tiny OPT}}= 0.55
~V_{\mbox{\scriptsize oc}}$.  By considering $V_{\mbox{\scriptsize
s}}=1/40$~ Volts, i.e. idealizing $u\alpha=1$ , $m/\gamma=0$, and
$V_{\mbox{\scriptsize oc}}=1$, we arrive at an upper bound of
FF=0.88 and $V_{\mbox{\scriptsize \tiny OPT}}= 0.91
~V_{\mbox{\scriptsize oc}}$, as shown in Fig.~\ref{fig-ff}(a).
For $u\alpha=0.7$ we obtain slightly smaller FF.
Fig.~\ref{fig-ff}(b) allows one to estimate one among
$V_{\mbox{\scriptsize s}}$, $V_{\mbox{\scriptsize oc}}$ and $FF$ from
the other two and will thus find practical application.  We estimated
the error bars for the experimental points from the regression fits
used to arrive at the values of $V_{\mbox{\scriptsize s}}$.  Devices
that we constructed locally had values of FF within these bounds.

Finally, our findings explain the very large fill factors of dye solar
cells.  The recombination current becomes insignificant as soon as the
voltage drops to $V=V_{\mbox{\scriptsize oc}}-V_{\mbox{\scriptsize
s}}$ (Eqs.~\ref{eq-jr}, \ref{eq-main}).  If $V_{\mbox{\scriptsize s}}
\ll V_{\mbox{\scriptsize oc}}$ (as in fact happens), then the
photocurrent jumps from zero to close to its short circuit value even
if the voltage only drops slightly (i.e., by $V_{\mbox{\scriptsize
s}}$). Notice from Fig.~\ref{fig-ff} that increases in
$V_{\mbox{\scriptsize oc}}$ --- e.g., due to greater radiant power ---
 should indeed lead to higher FF if
$V_{\mbox{\scriptsize s}}$ varies much less than $V_{\mbox{\scriptsize
oc}}$, as in fact occurs.

\section{Discussion and Conclusion}

We briefly comment on conversion efficiency.  Since we cannot expect
to change $V_{\mbox{\scriptsize s}}$ significantly, further increases
in efficiency will depend mainly on increasing the open circuit
photovoltage, which in turn also requires reducing losses due to
charge recombination at the nanoporous interface --- the main
challenge indeed.  Moreover, we know that $ c \sim n^{1/\gamma}, $ where
$\gamma $ has the negative value $\gamma \approx -6$. In other
words, we find evidence of a fractional power law or scaling exponent,
indicating self-affine behavior.  We hypothesize that the negative
value arises due to the fact that larger $n$ leads to greater
recombination, which consumes the oxidized species.  Localization
effects and and transport properties play an important role in this
context~\cite{nelson}.

In summary, our theoretical results appear to account well for the
observed behavior of real dye sensitized solar cells and seem to
provide new insights into their functioning.  Among the important
results reported here, we note that Fig.~\ref{fig-ff} allows one to
calculate FF knowing $V_{\mbox{\scriptsize oc}}/V_{\mbox{\scriptsize
s}}$ or vice versa. Moreover, knowing either one or the other, one can
readily obtain the entire photocurrent-voltage curve, via
Eq. \ref{eq-main}.  From just 3 points in the photocurrent-voltage
curve, one can reconstruct the entire curve.
The findings reported here
may thus allow further advances and eventually lead to technological
innovations.

We thank BNB, CAPES and CNPq for research funding and I. M. Gléria,
A. S. Gonçalves, M. L. Lyra, M. R. Meneghetti, S. M. P. Meneghetti,
F. A. B. F.  de Moura, A. F. Nogueira and L. S. Roman, E. C. Silva for
discussion.  GMV thanks S. B. Manamohanan for discussions in 1989.

\end{document}